\documentclass[11pt,twoside]{article}
\usepackage{asp2004}
\usepackage{psfig}
\usepackage{epsf}
\usepackage{graphics}
\usepackage{lscape}
\def\edcomment#1{\iffalse\marginpar{\raggedright\sl#1\/}\else\relax\fi}
\reversemarginpar

\begin{document}
\title{Gas accretion in galactic disks}
 \author{Thijs van der Hulst}
\affil{Kapteyn Astronomical Institute, P.O. Box 800, NL-9700 AV Groningen, 
the Netherlands}
\author{Renzo Sancisi}
\affil{INAF-Osservatorio Astronomico, Via Ranzani 1, I-40127 Bologna, Italy \\
Kapteyn Astronomical Institute, Groningen, the Netherlands}

\begin{abstract}
  
Evidence for the accretion of material in spiral galaxies has grown 
over the past years and clear signatures can be found in 
H\kern0.1em{\scriptsize I} observations of galaxies.  
We describe here new detailed and sensitive H\kern0.1em{\scriptsize I} 
synthesis observations of a few nearby galaxies (NGC~3359, NGC~4565 and 
NGC~6946) which show that 
indeed accretion of small amounts of gas is taking place.
These should be regarded as examples illustrating a general phenomenon 
of gas infall in galaxies. Such accretion may also be at the origin 
of the gaseous halos which are being found around spirals. 
Probably it is  the same kind of phenomenon of material infall as 
observed in the stellar streams in the halo and outer parts 
of our galaxy and M~31

\end{abstract}
\thispagestyle{plain}

\section{Introduction}

In this paper we present new evidence bearing on the formation of disks 
and halos of spiral galaxies through the accretion of small companions. 
This process is generally indicated as the nurture of galaxies. Evidence 
in support of it comes from various directions. The asymmetric shape  
of stellar disks (Zaritsky 1995 and Zaritsky \& Rix 1997) 
and the morphological and kinematic lopsidedness
observed in the H\kern0.1em{\scriptsize I} density distributions and 
velocity fields of spirals (Verheijen 1997, Swaters et al. 1999) may 
have originated from recent minor mergers. Furthermore, there is an 
increasingly large number of galaxies which in H\kern0.1em{\scriptsize I} 
show either peculiar features or clear signs of interactions with small 
companions (Sancisi 1999a and b). This indicates that galaxies often 
are in an environment where material for accretion is available.

Cold extra-planar gas has been found in several spiral galaxies. 
The best examples are those of NGC 891, NGC 2403 and UGC 7321  
(cf. Oosterloo et al., Fraternali et al., Matthews, this volume). 
The origin of this gas is not known. 
It has been suggested that it may, at least partly, be the product of 
galactic fountains. But some of its structural properties suggest that 
it may have originated from minor mergers. 

Recently, clear evidence that accretion events play an important
role has come from studies of the distribution and kinematics of stars
in the Milky Way halo. The discovery of the Sagittarius dwarf
galaxy (Ibata et al. 1994) is regarded as proof that accretion is
still taking place at the present time. Since such minor merger
remnants retain information about their origin for a long time (Helmi
\& White 2000) studies of the distribution and kinematics of ``stellar
streams'' can in principle be used to trace the merger history of the
Milky Way (Helmi \& de Zeeuw 2001). Such ``stellar streams'' are not
only seen in the Milky Way, but have also been discovered in the Local
Group galaxy M~31 (Ibata et al. 2001, Ferguson et al. 2002,
McConnachie et al. 2003). The substructure in the halo of M~31 is
another piece of clear evidence that minor mergers still take place.

Such events are difficult to trace in more distant galaxies,
where we can not observe individual stars. Other means are needed 
for detecting the signature of accretion. In this respect, the use of 
H\kern0.1em{\scriptsize I} is very powerful as it can image 
interactions very effectively by studying the H\kern0.1em{\scriptsize I}
distributions and kinematics. The latter is particularly useful for modelling.
Examples can be found in Sancisi (1999a). 
The improved sensitivity of modern synthesis radio telescopes brings 
within reach the detection of faint H\kern0.1em{\scriptsize I} 
signatures of accretion events and we expect that new observations of 
nearby and also more distant galaxies will reveal these in the coming 
decade. To further illustrate this point we here present a few examples 
of such signatures: NGC~3359, NGC~4565 and NGC~6946.

\begin{figure}[!ht] 
\plotfiddle{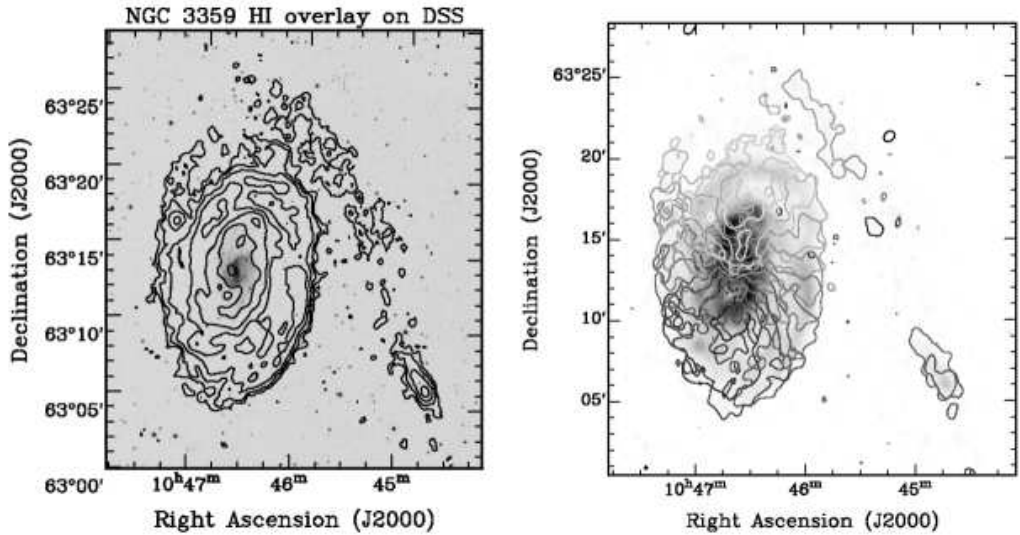}{8.0cm}{0}{44}{44}{-210}{-10}
\caption{The left panel shows the distribution of NGC~3359 at a resolution 
of 30$^{\prime\prime}$ superposed on the digital sky survey 
image. Contours are 0.1, 0.2, 0.4, 0.8, 1.6, 3.0, and 5.0 
$\times 10^{21}$ cm$^{-2}$. The right panel shows 
outer contours of the H\kern0.1em{\scriptsize I} emission in individual
channels superposed on the total H\kern0.1em{\scriptsize I} emission in 
NGC~3359. The low velocities are dark grey, the high velocities light grey. }
\end{figure}

\section{The observations}

All three galaxies have been observed recently with the Westerbork
Synthesis Radio Telescope (WSRT) using the new front-end and
correlator providing a much improved sensitivity. 
We will discuss each case individually below. Details of the observations 
and some of these results have already been reported by van der Hulst \& 
Sancisi (2004).
 
\subsection{NGC~3359}

NGC~3359 is a nearby barred spiral galaxy (Hubble type SBc) which
has been observed in H\kern0.1em{\scriptsize I} by Broeils (1992) as
part of a study of the mass distribution of a sample of nearby spiral
galaxies. It has a total mass of $1.2 \times 10^{11}$ M$_{\odot}$ and
an H\kern0.1em{\scriptsize I} mass of $7.5 \times 10^{9}$ M$_{\odot}$
(Broeils \& Rhee, 1997, adjusted for a Hubble constant of 72
km/s/Mpc).  It has well developed spiral structure both in the optical
and in H\kern0.1em{\scriptsize I}.  
Kamphuis \& Sancisi (1994, see
also Sancisi 1999a) pointed out the presence of an
H\kern0.1em{\scriptsize I} companion which appears distorted and has 
a long tail which may
connect to the H\kern0.1em{\scriptsize I} disk of NGC~3359. This
observation already indicated the  possible accretion
of gas by a large galaxy. Our new, more sensitive observations (rms
noise of 0.85 mJy/beam for velocity and spatial resolutions of 10 km s$^{-1}$ 
and 30$^{\prime\prime}$ respectively) are shown in Figure 1 (left panel)
and convincingly display an H\kern0.1em{\scriptsize I} connection between
the distorted H\kern0.1em{\scriptsize I} companion and the main
galaxy. The mass of the H\kern0.1em{\scriptsize I} companion is $1.8
\times 10^{8}$ M$_{\odot}$ or 2.4\% of the H\kern0.1em{\scriptsize I} 
mass of NGC~3359. 
The H\kern0.1em{\scriptsize I} distribution of the companion 
is clearly distorted and shows a tail pointing towards and 
connecting with the outer spiral structure of NGC~3359. 
No clear optical counterpart of the
H\kern0.1em{\scriptsize I} companion has yet been found.
\begin{figure}[!ht]
\plotfiddle{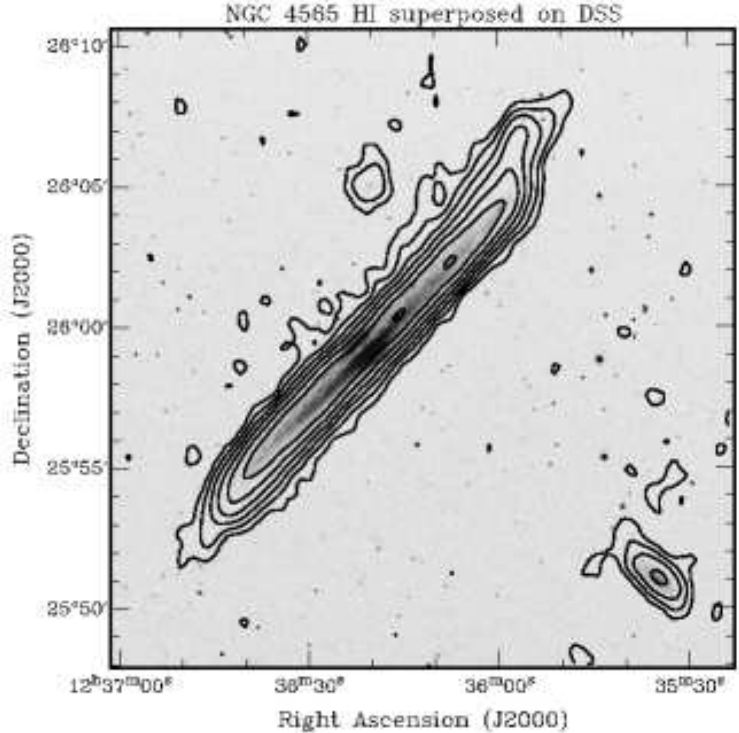}{10.5cm}{0}{60}{60}{-175}{-23}
\caption{H\kern0.1em{\scriptsize I} distribution of NGC~4565 at a resolution 
of 30$^{\prime\prime}$ superposed on the digital sky survey 
image. Contours are 0.2, 0.4, 0,8, 1.6, 3.2, and 6.4 
$\times 10^{21}$ cm$^{-2}$.}
\end{figure}

The velocity structure of the H\kern0.1em{\scriptsize I} companion and the 
connecting H\kern0.1em{\scriptsize I} fits in very well with the regular 
velocity field of NGC~3359. This is
shown in the right panel of Figure 1 
where we display the emission in the individual
channels superposed on the total H\kern0.1em{\scriptsize I} image of 
NGC~3359.  
Contours of
different shades of grey (low velocities are dark, high velocities are
light) denote the outer edge of the H\kern0.1em{\scriptsize I} emission in 
each of the velocity channels and thus display the basic kinematics of the 
H\kern0.1em{\scriptsize I}
without any further analysis of individual velocity profiles. The
regularity of the velocities suggests that the process has been going
on slowly for at least one rotational period which is of the order of
1.7 Gy.
\begin{figure}[!ht] 
\plotone{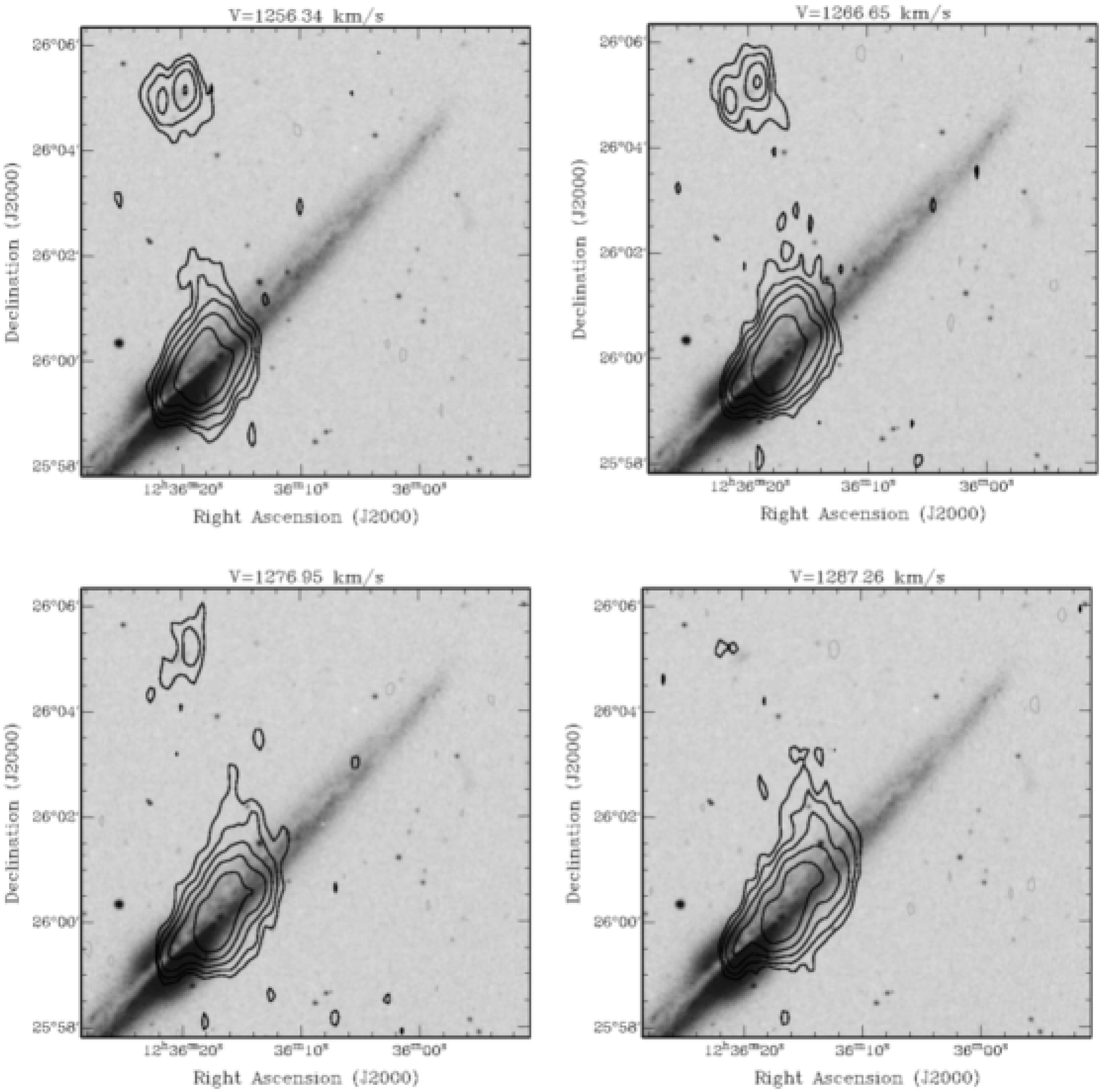}
\caption{H\kern0.1em{\scriptsize I} emission at four velocities between
1256 and 1287 km s$^{-1}$ superposed on the DSS image of NGC~4565 at a 
resolution of  $13^{\prime\prime} \times 33^{\prime\prime}$. 
These channels clearly show the interaction between the companion and
NGC~4565. 
Contours are -2.0 -1.0 1.0 2.0 4.0 8.0 160. 32 0. 64.0 120.0 
mJy/beam}
\end{figure}

\begin{figure}[!ht] 
\plotone{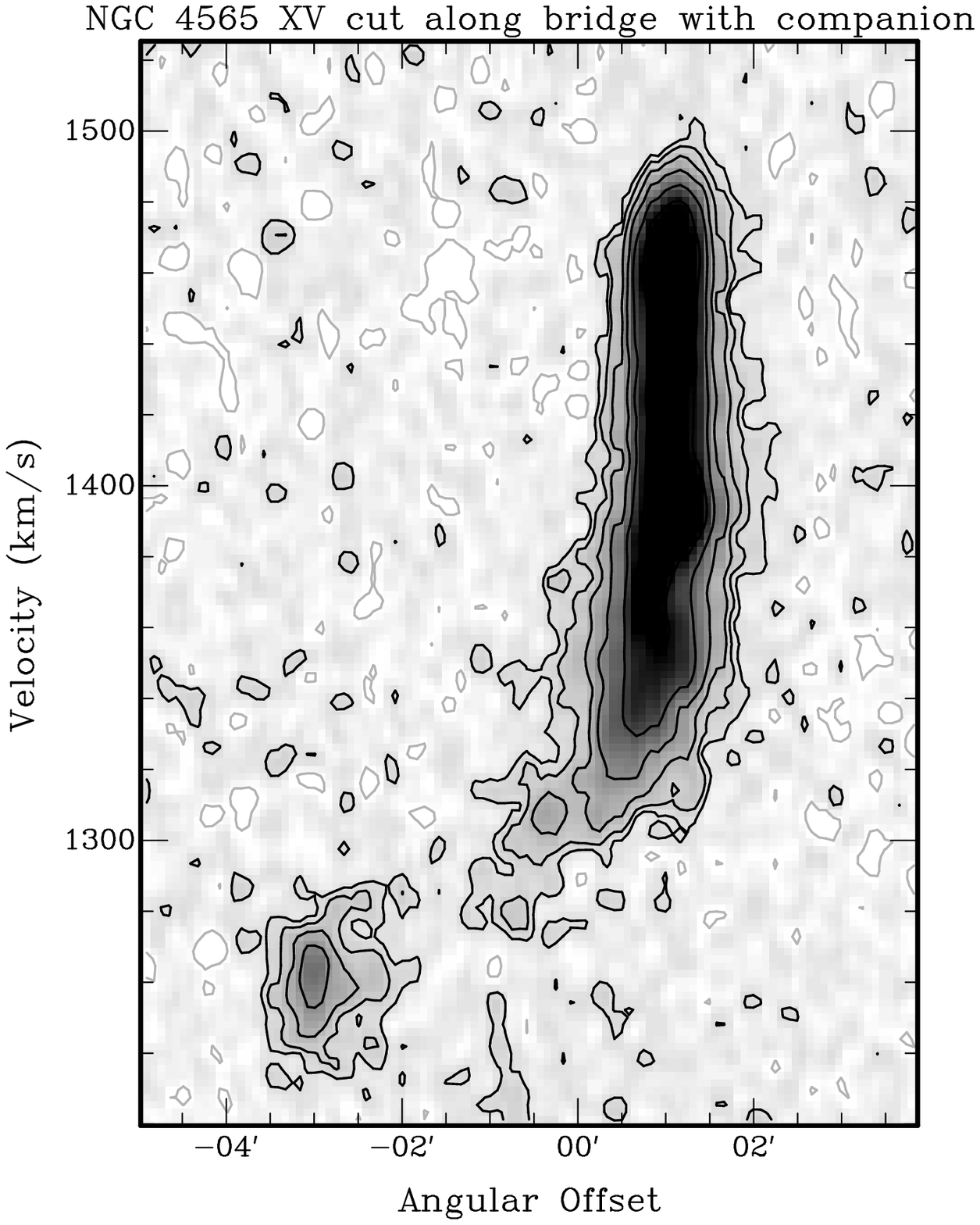}
\caption{Position velocity diagram parallel to the minor axis of NGC~4565
across F378-0021557 and the disk of NGC~4565.}
\end{figure}

\subsection{NGC~4565}

NGC~4565 is a large edge-on galaxy of Hubble type Sb which was first
observed in H\kern0.1em{\scriptsize I} by Sancisi (1976) in an early
search for galaxies with warped H\kern0.1em{\scriptsize I}
disks. Rupen (1991) observed NGC~4565 with much higher resolution and
presented a detailed study of the kinematics and the warp.  NGC~4565
has a small optical companion 6$^{\prime}$ to the north of the center, 
F378-0021557, which has $7.4 \times 10^{7}$ M$_{\odot}$
of H\kern0.1em{\scriptsize I} compared to an H\kern0.1em{\scriptsize
I} mass of $2.0 \times 10^{10}$ M$_{\odot}$ for NGC~4565 (using
a distance of 17 Mpc). An H\kern0.1em{\scriptsize I}
detection of this companion called NGC 4565A has also been reported 
by Rupen (1991). Another H\kern0.1em{\scriptsize I}
companion, NGC~4562, somewhat larger in H\kern0.1em{\scriptsize I}
($2.5 \times 10^{8}$ M$_{\odot}$) and brighter optically can be found
15$^{\prime}$ to the south-west of the center of NGC~4565. The
H\kern0.1em{\scriptsize I} distribution, derived from a new sensitive
WSRT observation by Dahlem (priv. comm.), is shown in Figure 2
superposed on the DSS. The asymmetric warp is clearly visible.

Inspection of individual channel maps brings to light that in addition
to the warp the H\kern0.1em{\scriptsize I} distribution shows
additional, low surface brightness emission to the north of the
center, in the direction of the faint companion F378-0021557.  The
H\kern0.1em{\scriptsize I} emission in the velocity range from 1250 to
1290 km s$^{-1}$ (close to the velocity of F378-0021557 and to the
systemic velocity, 1230 km s$^{-1}$, of NGC~4565) clearly shows
distortions above the plane pointing towards the companion.  This is
best seen in Figure 3 which shows four channel maps chosen at velocities
in this range. In these maps one can clearly see the
H\kern0.1em{\scriptsize I} layer bending towards
F378-0021557, indicating a connection between F378-0021557 and a strong
disturbance in the H\kern0.1em{\scriptsize I} disk of NGC~4565.  This
disturbance is not associated
with the warp.  However, this bending of the H\kern0.1em{\scriptsize
I} layer, undoubtedly caused by the companion, is remarkably similar
tot the outer H\kern0.1em{\scriptsize I} warping. One wonders whether
the type of interaction we are witnessing here is the mechanism also
responsible for the creation of a warp.  This interaction
between NGC~4565 and its companion will eventually lead to a merger 
of the latter with NGC~4565.

To further elucidate the connection between F378-0021557 and NGC~4565
we show a position-velocity diagram along the H\kern0.1em{\scriptsize I}
connection in figure 4. This position-velocity cut has been taken parallel
to the minor axis of NGC~4565 through F378-0021557 and clearly shows the
velocity continuity of the H\kern0.1em{\scriptsize I} disturbance in
the disk of NGC~4565 and F378-0021557.

\subsection{NGC~6946}

NGC~6946 is a bright, nearby spiral galaxy of Hubble type Scd
which has been studied in H\kern0.1em{\scriptsize I} numerous times
(Rogstad et al. 1973, Tacconi \& Young 1986, Kamphuis 1993). It was in
this galaxy that Kamphuis and Sancisi (1993) found the first evidence
for an anomalous velocity H\kern0.1em{\scriptsize I} component which
they associated with outflow of gas from the disk into the halo as a
result of stellar winds and supernova explosions. Evidence for such a
component is now being found in more galaxies as discussed by
Fraternali et al. (2001, 2002, and also this volume). A detailed study
of the anomalous H\kern0.1em{\scriptsize I} and the structure in the
H\kern0.1em{\scriptsize I} disk is being carried out by Boomsma et
al. (this volume) on the basis of very sensitive observations with the
WSRT (rms of 0.5 mJy/beam for spatial and velocity resolutions of
60$^{\prime\prime}$ and 5 km s$^{-1}$ respectively).

\begin{figure}[!ht]
\plotfiddle{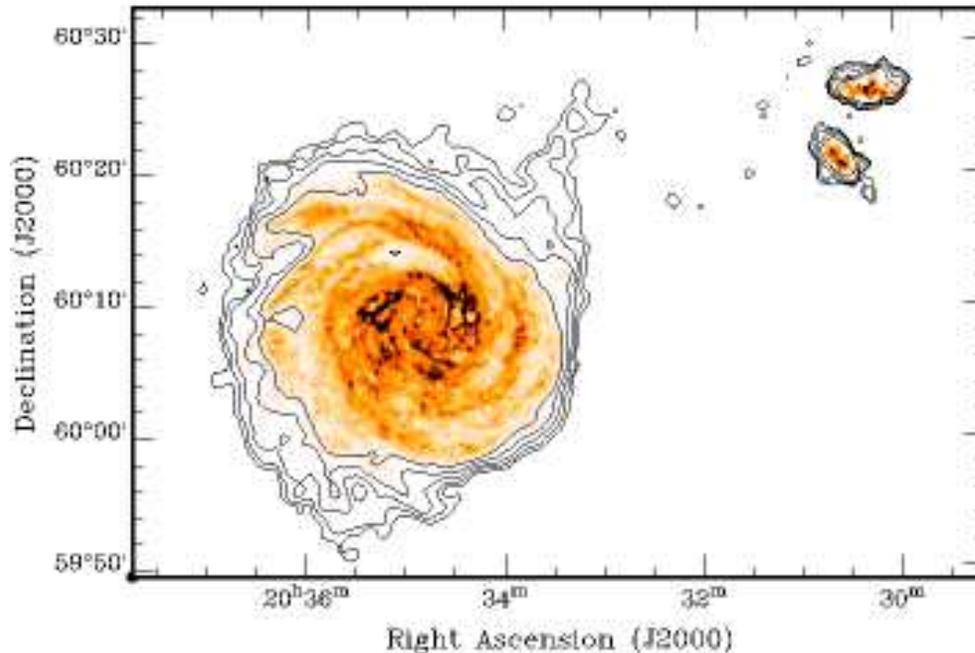}{8.35cm}{0}{72}{72}{-210}{-70}  
\caption{H\kern0.1em{\scriptsize I} distribution in NGC~6946 at a resolution 
of 30$^{\prime\prime}$ (greyscale) and
60$^{\prime\prime}$ (contours, see also Boomsma et al. this volume). 
Contours are 1.3, 2.6, 5.2, 10.4, and 20.8  $\times 10^{19}$ cm$^{-2}$. }
\end{figure}

Here we use a low resolution (60$^{\prime\prime}$) version of these
data. Figure 6 shows a total H\kern0.1em{\scriptsize I} image of
NGC~6946 down to column density levels of $1.3 \times 10^{19}$
cm$^{-2}$. To the west two small companion galaxies can be seen. The
most intruiging feature is the faint whisp to the north-west of the
H\kern0.1em{\scriptsize I} disk of NGC~6946.  This faint
H\kern0.1em{\scriptsize I} extension can only be brought out at this
resolution and appears to form a faint H\kern0.1em{\scriptsize I}
filament which blends smoothly (also kinematically) with the
H\kern0.1em{\scriptsize I} disk of NGC~6946 at a position some
11$^{\prime}$ (or 19 kpc) south of the tip of the filament. There is
no detected connection with the two companion galaxies farther to the
west. The spatial and velocity structure of the object are so regular,
yet only connected to the main H\kern0.1em{\scriptsize I} disk at one
side that we prefer an explanation in terms of a tidally stretched,
infalling H\kern0.1em{\scriptsize I} object. So this looks like yet 
another example of accretion of small amounts of gas onto a large 
H\kern0.1em{\scriptsize I} disk.

Similar examples are perhaps the filament discovered in NGC~2403 
(Fraternali et al. 2001, 2002 and also this volume), a long 
H\kern0.1em{\scriptsize
I} filament in M~33 (van der Hulst, unpublished) and the extra-planar
filaments in the northern part of the H\kern0.1em{\scriptsize I} halo
of NGC~891 (Fraternali et al., this volume).

\section{Concluding remarks}

We have shown three cases with strong evidence for the accretion of
small amounts of H\kern0.1em{\scriptsize I}. These are certainly not unique. 
There are several more cases known (Sancisi 1999a and b). Furthermore, there 
are cases with such faint features that can only be seen in sensitive 
H\kern0.1em{\scriptsize I} observations as the H\kern0.1em{\scriptsize I} 
masses involved are rather modest. We therefore expect that with the
increased sensitivity of modern synthesis radio telescopes, more
examples will be discovered in the coming decade. There probably is a
range of H\kern0.1em{\scriptsize I} masses for these accretion events 
as is already apparent from the six cases mentioned here: NGC~891, NGC~2403, 
NGC~3359, NGC~4565, NGC~6946 and M~33.

What is the effect of accretion on the main galaxy? 
This can influence local star formation in the disks and starbursts.
For instance, there may very well be a connection with the star formation 
activity in the disks of galaxies such as NGC~6946 and NGC~2403.
The infall of gas may be at the origin of the extra-planar gas and gaseous 
halos recently discovered in spiral galaxies.
Also, it may affect the structure of galactic disks in the outer parts 
and possibly also contribute to the formation of the outer layers and 
of warps in particular. 

It is quite clear that future sensitive and detailed studies of the 
H\kern0.1em{\scriptsize I} in nearby galaxies will provide a more 
complete census of the phenomena discussed in this paper and enable 
us to address these issues further and obtain more definitive answers. 
In particular, it should be possible to obtain estimates of the mean 
gas accretion rate in galaxies.





\begin{thebibliography}

\bibitem []{broeilsPhD}
 Broeils, A.~H., 1992, PhD thesis, University of Groningen
\bibitem []{broeils97}
 Broeils, A.~H., Rhee, M.~-H., 1997, \aap, 324, 877
\bibitem []{ferguson02}
 Ferguson, A.~M.~N., Irwin, M.~J., Ibata, R.~A., Lewis, G.~F., 
        \& Tanvir, N.~R.\ 2002, \aj, 124, 1452 
\bibitem []{fraternali01}
 Fraternali, F., Oosterloo, T., Sancisi, R., \& 
        van Moorsel, G.\ 2001, \apjl, 562, L47 	
\bibitem []{fraternali02}
 Fraternali, F., van Moorsel, G., Sancisi, R., \& 
        Oosterloo, T.\ 2002, \aj, 123, 3124 
\bibitem []{helmi00}
 Helmi, A.~\& Tim de Zeeuw, P.\ 2000, \mnras, 319, 657 
\bibitem []{helmi01}
 Helmi, A.~\& White, S.~D.~M.\ 2001, \mnras, 323, 529 
\bibitem []{ibata01}
 Ibata, R., Irwin, M., Lewis, G., Ferguson, A.~M.~N., \& 
        Tanvir, N.\ 2001, Nature, 412, 49 
\bibitem []{ibata94}
 Ibata, R.~A., Gilmore, G., \& Irwin, M.~J.\ 1994, Nature, 370, 194 
\bibitem []{kamphuisPhD}
 Kamphuis, J.\ 1993, PhD thesis, University of Groningen
\bibitem []{kamphuis93}
 Kamphuis, J.~\& Sancisi, R.\ 1993, \aap, 273, L31 
\bibitem []{kamphuis94}
 Kamphuis, J.~\& Sancisi, R.\ 1994, Panchromatic View of 
        Galaxies.~Their Evolutionary Puzzle, eds. G. Hensler, C. Theis 
	and J.S. Gallagher, Editions Frontiers, p. 317  
\bibitem []{mcconnachie03}
 McConnachie, A.~W., Irwin, M.~J., Ibata, R.~A., Ferguson, 
        A.~M.~N., Lewis, G.~F., \& Tanvir, N.\ 2003, \mnras, 343, 1335 
\bibitem []{rogstad73}
 Rogstad, D.~H., Shostak, G.~S., \& Rots, A.~H.\ 
        1973, \aap, 22, 111 
\bibitem []{rupen91}
 Rupen, M.~P.\ 1991, \aj, 102, 48 
\bibitem []{sancisi76}
 Sancisi, R.\ 1976, \aap, 53, 159 
\bibitem []{sancisi99a}
 Sancisi, R.\ 1999a, IAU Symp.~186: Galaxy Interactions at Low 
        and High Redshift, eds. J.E. Barnes and D. B. Sanders, p. 71 
\bibitem []{sancisi99b}
 Sancisi, R.\ 1999b, \apss, 269, 59  
\bibitem []{tacconi86}
 Tacconi, L.~J.~\& Young, J.~S.\ 1986, \apj, 308, 600 
\bibitem []{vdhulst02}
 van der Hulst, J.~M.~\& Sancisi, S. 2004, IAU Symp.~217: Recycling 
 intergalactic  and interstellar matter, eds. P.-A. Duc, J. Braine and E.
 Brinks, p. 122
\bibitem []{zaritsky95}
 Zaritsky, D.\ 1995, \apjl, 448, L17 
\bibitem []{zaritsky97}
 Zaritsky, D.~\& Rix, H.\ 1997, \apj, 477, 118 

\end{thebibliography}
\end{document}